\documentclass[conference]{IEEEtran}

\usepackage[pdftex]{graphicx}
\graphicspath{{./fig/}}
\DeclareGraphicsExtensions{.pdf}
\usepackage[caption=false,font=normalsize,labelfont=sf,textfont=sf]{subfig}

\usepackage{amsmath}

\usepackage{algorithmic}

\usepackage{booktabs}

\usepackage{fixltx2e}

\usepackage{url}
\usepackage{cite}

\begin{document}
\title{Energy Scaling with Control Distance\\ in Complex Networks}
\author{\IEEEauthorblockN{Isaac Klickstein, Ishan Kafle, Sudarshan Bartaula, and Francesco Sorrentino}
\IEEEauthorblockA{Department of Mechanical Engineering\\University of New Mexico\\Email: iklick@unm.edu}
}
\maketitle
\begin{abstract}
  It has recently been shown that the expected energy requirements of a control action applied to a complex network scales exponentially with the number of nodes that are targeted.
  While the exponential scaling law provides an adequate prediction of the mean required energy, it has also been shown that the spread of energy values for a particular number of targets is large.
  Here, we explore more closely the effect distance between driver nodes and target nodes and the magnitude of self-regulation has on the energy of the control action.
  We find that the energy scaling law can be written to include information about the distance between driver nodes and target nodes to more accurately predict control energy.
\end{abstract}
\IEEEpeerreviewmaketitle
\section{Introduction}
The control of complex networks is an extremely active field \cite{ruths2014control,liu2016control,klickstein2017energy,yan2015spectrum,shirin2017optimal}.
Numerous applications can be found from regulating power grids, routing traffic on the internet, marketing on social media, synchronization of multi-agent systems, and many others.
Of particular interest recently is the specific control action that is minimal with respect to the control energy \cite{klickstein2017energy, yan2015spectrum, yan2012controlling, li2015minimum,klickstein2017locally}.\\
\indent
The nodes of the network are classified as driver nodes, target nodes, or neither.
We will call those nodes which receive a control signal directly \emph{driver nodes} \cite{liu2016control} and those nodes whose state values we prescribe at the end of the control action \emph{target nodes} \cite{iudice2015structural}. 
It has been shown recently that the expected amount of energy required for a particular control action on a complex network decreases exponentially with respect to the number of inputs (or driver nodes) \cite{yan2015spectrum} and increases exponentially with respect to the number of outputs (or target nodes) \cite{klickstein2017energy}.
These scaling relations describe very well the required amount of control energy when this quantity is averaged over many possible network realizations.
The variation seen in the energy requirement about the exponential mean value is directly related to the topology of the graph that describes the connections between vertices.
Other scaling relations have been obtained in \cite{wang2012optimizing,chen2016energy}, which focus on the case that one wants to target all the network nodes.
The main difference with the previous work is that here we exploit the recently developed framework for optimal target control of networks \cite{klickstein2017energy,shirin2017optimal} to provide new insight into how the selection of both driver and target nodes over a network affects the control energy.
When trying to increase the controllability of a complex network, it is often the addition of edges \cite{wang2012optimizing,chen2016energy} that provides the most improvement.\\
\indent
In section II we briefly provide background on graph theory and optimal control of complex networks.
In section III we present results on both canonical graphs (paths and stars) as well as large Erdos-Renyi graphs \cite{newman2010networks}.
Finally, we summarize how the energy scales with respect to the graph properties investigated in section IV.
\section{Background}
\begin{figure*}[!t]
  \centering
  \includegraphics[scale=1]{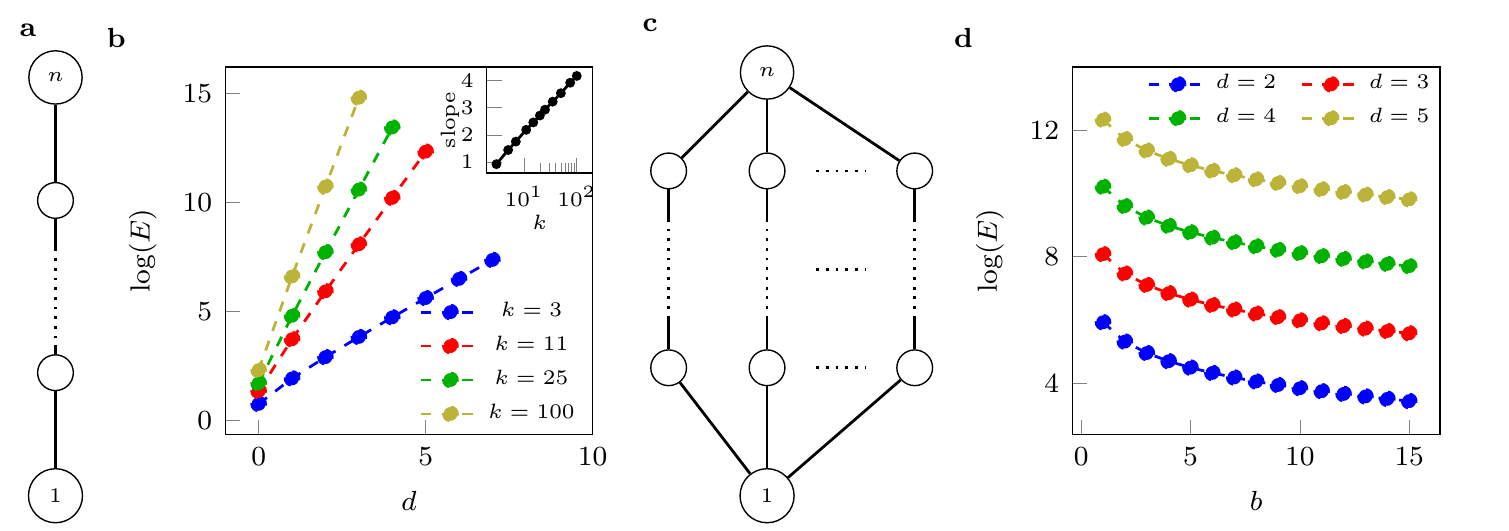}
  \caption{
    Energy requirements for two types of graphs.
    \textbf{a} A path graph of length $d = n-1$.
    \textbf{b} The energy required when the driver node is at one end of the path and the target node is chosen to be distance $d$ away for different values of self-regulation parameter $k$.
    As $k$ increases, the curves become steeper, while remaining linear.
    The inset shows how the slope of these curves increase with increasing regulation parameter $k$.
    \textbf{c} A balloon graph of length $d$ and number of branches $b$ so that $n = b(d-1)+2$.
    The driver node is placed at one of the hubs and the target is chosen as the other hub.
    \textbf{d} The energy to control one hub with the other hub of the balloon graph for varying $b$ and $d$.
    As the number of branches increase, while holding the distance between driver and target constant, the energy required decreases by a few orders of magnitude.
  }
  \label{fig:path}
\end{figure*}
\begin{figure}
  \centering
  \includegraphics[scale=1]{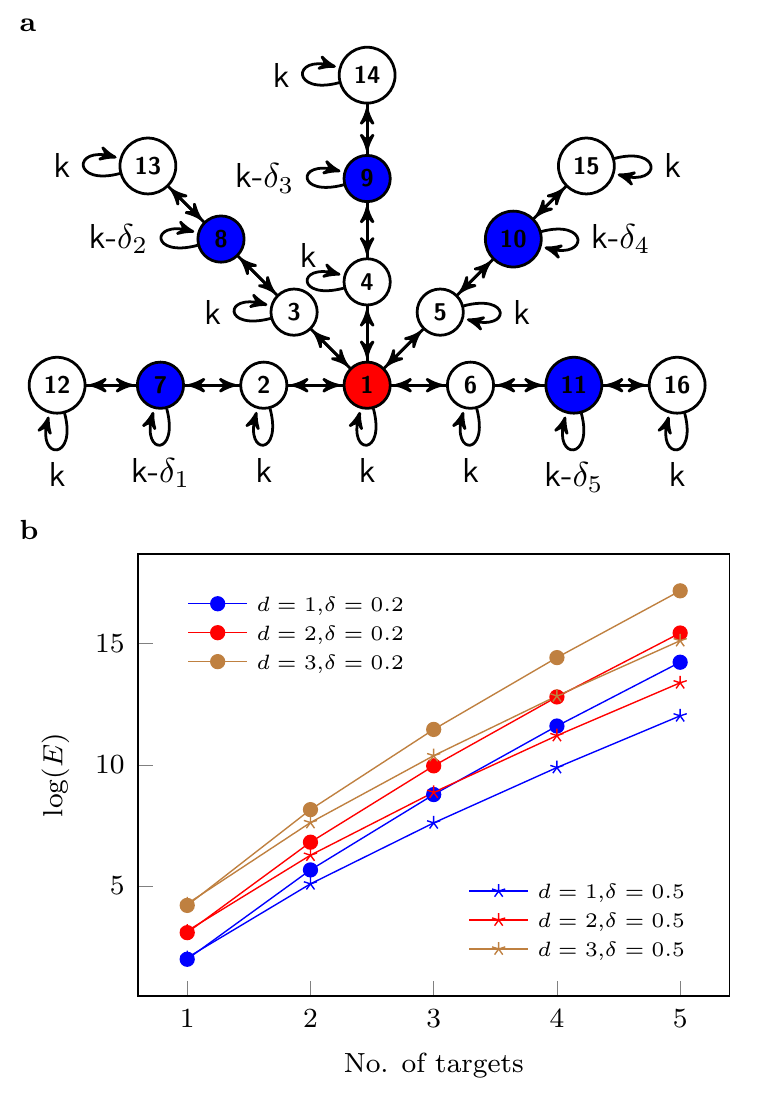}
  \caption{
    Analyzing the energy requirements for a star graph while increasing the number of targets.
    \textbf{a} A star graph with branches of length $3$.
    We emphasize that we analyzed the case where each target added to the set of target nodes is at the same distance from the single driver node located at the hub.
    \textbf{b} The exponential scaling with respect to the cardinality of the target set $p$ is expected.
    On the other hand, we see that the rate of the exponential increase when adding target nodes at a given distance is approximately independent of the distance (see Table \ref{tab:star}).
  }
  \label{fig:star}
\end{figure}
A graph $\mathcal{G} = (\mathcal{V},\mathcal{E})$ consists of a set of $n$ nodes $\mathcal{V} = \{v_i | i=1,\ldots,n\}$ and $\ell$ edges $\mathcal{E} \subset \mathcal{V} \times \mathcal{V}$ where $(v_i,v_j) \in \mathcal{E}$ if node $v_j$ receives information from node $v_i$.
Our work focuses on undirected graphs, that is $(v_i,v_j) \in \mathcal{E}$ implies $(v_j,v_i) \in \mathcal{E}$.
A graph can be represented as an $n$ by $n$ adjancency matrix $A$ which has elements $A_{ij} = A_{ji} = 1$ if the pair $(v_i,v_j) \in \mathcal{E}$ and $A_{ij} = A_{ji} = 0$ otherwise for $i \neq j$.
We assume there is a self-loop at every node, that is, $(v_i,v_i) \in \mathcal{E}$ $i = 1,\ldots,n$, with weight $-k_i$ so the diagonal values of the adjacency matrix are $A_{ii} = -k_i$.
Given a pair of nodes $(v_i,v_j)$, we define the \textit{distance} between them $d_{ij}$ as the length of the shortest path from $v_i$ to $v_j$.
As the graph $\mathcal{G}$ is undirected $d_{ij} = d_{ji}$.\\
\indent
Many dynamical systems can be described by graphs where the edges of the graph represent transfer of power, traffic, influence, or information.
We consider linear dynamics where the $n$ by $n$ state matrix $A$ is the adjacency matrix of the graph and the $n$ by $m$ control matrix $B$ describes where in the network the control inputs $u_k(t)$, $k = 1, \ldots, m$ are attached.
The $p$ by $n$ output matrix $C$ selects the target nodes.
\begin{equation}\label{eq:sys}
  \begin{aligned}
    \dot{\textbf{x}}(t) &= A \textbf{x}(t) + B \textbf{u}(t), && \textbf{x}(t_0) = \textbf{x}_0\\
    \textbf{y}(t) &= C \textbf{x}(t), && \textbf{y}(t_f) = \textbf{y}_f
  \end{aligned}
\end{equation}
The initial condition of the system $\textbf{x}_0$ is known and the final condition for the target nodes $\textbf{y}_f$ is prescribed.
To enforce the graph nature of this problem, we restrict $B$ ($C$) to have $n$-dimensional, independent versors as columns (rows).
If entry $B_{ij} = 1$, then node $i$ is a driver node, and if entry $C_{ji} = 1$, then node $i$ is a target node.\\
\indent
In the following statements, we will assume the triplet $(A,B,C)$ is \emph{output controllable} \cite{antoulas2005approximation}, that is the rank of the matrix $\text{rank}(K) = \text{rank}\left[CB|CAB|\ldots|CA^{n-1}N\right] = p$.
The control input $\textbf{u}^*(t)$ that is minimal with respect to energy $E = \int_{t_0}^{t_f} ||\textbf{u}(t)||^2 dt$ that drives the system in Eq. \eqref{eq:sys} from the initial condition $\textbf{x}_0$ to the prescribed final condition for the targets $\textbf{y}_f$ is known from optimal control theory \cite{klickstein2017energy},
\begin{equation}\label{eq:u}
  \textbf{u}(t) = B^T e^{A^T(t_f-t)} C^T \left(CW(t_f)C^T\right)^{-1} \textbf{g}
\end{equation}
The symmetric semi-positive definite matrix $W(t)$ is the controllability Gramian which solves the linear ODE $\dot{W} = AW(t) + W(t)A^T + BB^T$.
The vector $\textbf{g}=\left(\textbf{y}_f - C e^{A(t_f-t_0)} \textbf{x}_0 \right)$ is the difference between the prescribed final conditions and the zero-input evolution of the system when $t=t_f$.
If the matrix $A$ is Hurwitz, then the ODE that describes the evolution of $W(t)$ is globally, asymptotically stable with fixed point $W^*$ that solves the equation,
\begin{equation}\label{eq:lyap}
  O_n = AW^* + W^* A^T + BB^T
\end{equation}
Equation \eqref{eq:lyap} is an algebraic Lyapunov equation which can be solved efficiently.
The characteristic energy of the control action can be written succinctly in terms of the asymptotic controllability Gramian $W^*$.
\begin{equation}
  \begin{aligned}
      E^* &= \max\limits_{||\textbf{g}||=1} \lim\limits_{t_f \rightarrow \infty} \int_{t_0}^{t_f} ||\textbf{u}(t)||^2 dt\\
      &= \max\limits_{||\textbf{g}||=1} \textbf{g}^T \left(CW^*C^T\right)^{-1} \textbf{g} = \frac{1}{\mu^*_{\min}}
    \end{aligned}
  \end{equation}
  The positive scalar $\mu^*_{\min}$ is the smallest eigenvalue of the matrix $CWC^T$.
  Note that if $p = 1$, then $C$ has a single non-zero entry located at position $i$ and so $CW^*C^T = W^*_{ii} = \mu^*_{\min}$.
  The case when $m = 1$ and $p = 1$ is of particular interest to us as it represents the energetic contribution when controlling a single target node from a single driver node.\\
  \indent
  As we will assume that $A$ is Hurwitz, we choose the diagonal values of the adjacency matrix $k_i$ to ensure this property holds.
  By the Gerschgorin Disc Theorem, we choose $-k_i < -\max\{\text{deg}(i)\}$ where $\text{deg}(i)$ is the degree of node $i$ so that $k_i > \sum_{j=1,j\neq i}^n |A_{ij}|$.
  \\
  \indent
  From our studies of the control energy scaling of complex networks, we hypothesize that the particular contribution to the control energy any one target node provides is a function of three main quantities; (i) the \emph{distance} between the driver node and the target node, (ii) the \emph{redundancy of paths} between the driver node and the target node, and (iii) the cardinality of the set of target nodes.
  These three aspects will be investigated in the next section.
\section{Results}
\begin{figure}[!t]
  \centering
  \includegraphics[scale=1]{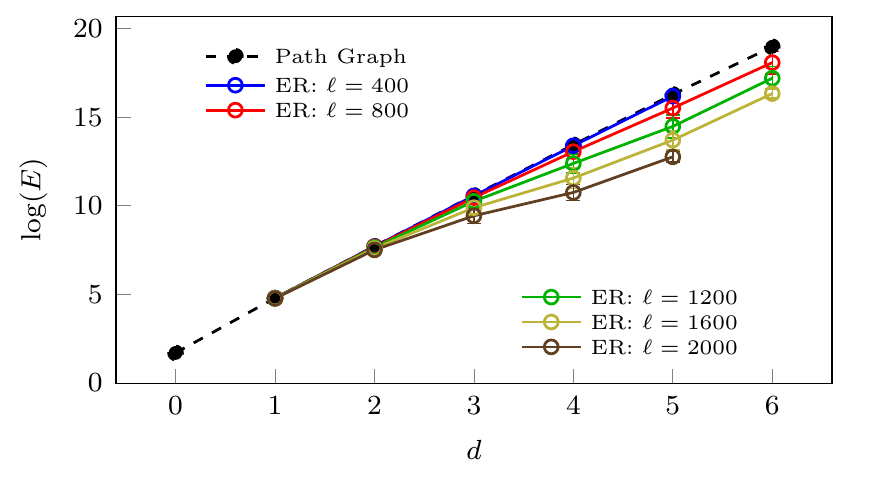}
  \caption{
    The energy requirements for the one driver and one target problem in Erdos-Renyi graphs of different density, $\ell/n$.
    There is at most one path of length $d = 1$ between two nodes and so the energy for pairs of nodes in any of the ER graphs of distance $1$ apart is equal.
    As the distance between pairs of nodes grows, the number of paths of that distance also grows and so, as was seen in the analysis of the balloon graphs, the energy decreases at most a few orders of magnitudes.
    Similar results were obtained for $k$-regular graphs and scale-free graphs.
  }
  \label{fig:ER}
\end{figure}
For the first property that determines the energy contribution of a target node, we examine the role that distance between driver nodes and target nodes plays.
We consider a path graph of length $n$ where the driver node is chosen to be at one end which we define to be node $1$.
A path graph is shown in Fig. \ref{fig:path}\textbf{a}.
The target node is chosen along the path so that if node $i$ is the target node, then the driver node and target node are distance $(i-1)$ apart.
The energy of the control action is $E^* = 1/W_{ii}^*$, and $\log(E^*)$ is shown in Fig. \ref{fig:path}\textbf{b}.
We perform the simulation for path graphs with increasing values of $k$ along the main diagonal of the adjacency matrix.
As $k$ grows, the curves in Fig. \ref{fig:path}\textbf{b} become steeper.
Interestingly, we see there is an exponential scaling of the energy with distance between driver node and target node so that $\log(E^*) \propto f(k) d$, where $f(k)$ represents the slope of the energy curve for a path graph with $-k$ along the diagonal of the adjacency matrix.
The inset of Fig. \ref{fig:path}\textbf{b} shows how the slope of these energy curves increases with respect to $k$ where we see linear dependence such that $f(k) \propto 2\log (k)$.
Combining these results,
\begin{equation}\label{eq:path}
  \log (E) \propto 2 \log (k) d
\end{equation}
While the energy scaling in Eq. \eqref{eq:path} provides a reasonable prediction for the path graph, when examining the energy requirements for the case of one driver node and one target node in a large complex network, it is far less successful.
In fact, given a set of pairs of driver nodes and target nodes in a complex network of all the same distance apart, the energy may vary across multiple orders of magnitude.
This is due, in the case of a complex network, to the redundancy of paths of a given length between pairs of nodes.\\
\indent
To investigate this phenomenon, we turn to a generalization of the path graph which we call a balloon graph as shown in Fig. \ref{fig:path}\textbf{c}.
A balloon graph consists of two hubs labeled $1$ and $n$ with $b$ vertex-disjoint branches that connect the hubs, each of length $d$ so that the whole graph consists of $n = b(d-1)+2$ nodes.
When $b = 1$, the balloon graph contracts to a path graph of length $d$.
The energy curves for different numbers of branches of different lengths is shown in Fig. \ref{fig:path}\textbf{d}.
We see that as the number of branches increases from one to three, the energy is reduced most drastically
As the number of branches continues to grow, the energy continues to reduce about linearly.
This decay provides a modification to Eq. \eqref{eq:path} by including a term that describes the number of vertex-disjoint paths of length $d$.
\begin{equation}\label{eq:branch}
  \log(E) \propto 2 \log(k) d - \alpha b
\end{equation}
The coefficient $\alpha$ is obtained by finding the slope of the linear portions of the curves in Fig. \ref{fig:path}\textbf{d} where we see $\alpha \approx \frac{1}{10}$ for all curves.
We have seen that for the case where we include additonal paths of length larger than $d$, the decrease in energy is much less substantial, that is, the energy  value is dominated by the length of the shortest path between driver node and target node.
On the other hand, including an additional path of length smaller than $d$ reduces the energy far more than seen in Fig. \ref{fig:path}\textbf{d} as we have reduced the distance between the pair of nodes.
This indicates that though all the paths between pairs of nodes may contribute to the control energy, the shortest path between them is the one whose contribution is dominant.\\
\begin{figure}[!t]
  \centering
  \includegraphics[scale=1]{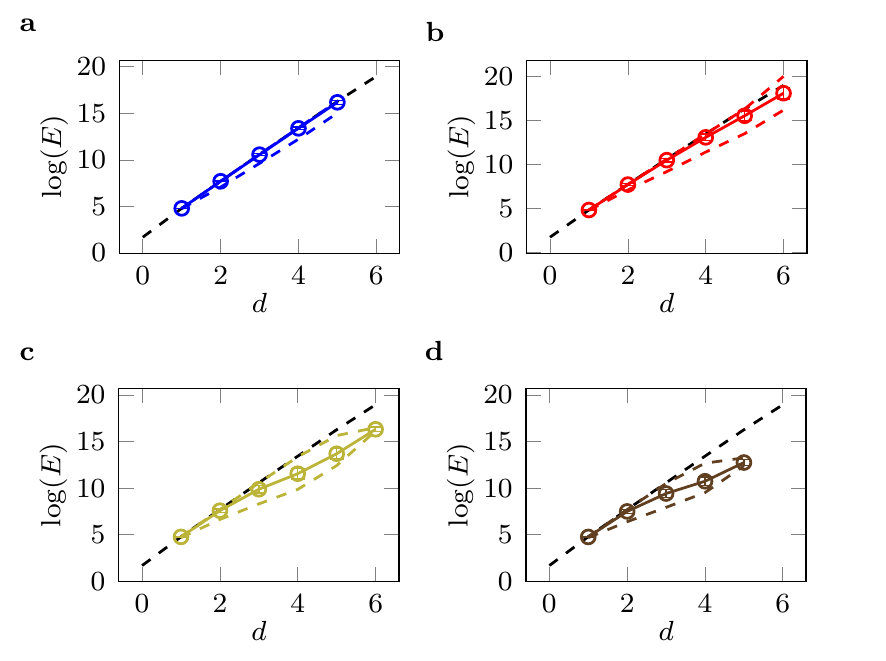}
  \caption{
    The spread of energy requirements for the one driver and one target case in Erdos-Renyi graphs of increasing density.
    \textbf{a} $m=400$, \textbf{b} $m = 800$, \textbf{c} $m = 1600$, \textbf{d} $m = 2000$.
    The top dashed line in each panel is the maximum value of $\log E^*$ among all pairs of nodes at distance $d$.
    The bottom dashed line in each panel is the minimum value of $\log E^*$ among all pairs of nodes at distance $d$.
    We see that the upper limit of the energy is bounded by a path graph with equal regulation parameter $k$ along the diagonal of the adjacency matrix.
    The lower limit of the energy is at most a few orders of magnitude less as was seen in the analysis of the balloon graph.
  }
  \label{fig:max}
\end{figure}
\indent
Here we extend our study to the case of mutliple target nodes.
As the shortest distance is the dominating factor in affecting the control energy, we choose target nodes that are all at the same distance from a given driver node.
We consider the star graph in Fig. \ref{fig:star}\textbf{a}, where the driver node is in red and all the nodes at distance 2 are in blue.
We can vary both the distance of the target nodes, as well as the number $p$ of target nodes at that given distance.
Without loss of generality we label the target nodes $1,2,...,p$.
In order to maintain controllability, we set the self loop weights at each target node, $k_i=k -i \delta$, $i = 1,\ldots,p$, so that the parameter $\delta$ measures the minimum difference between any two self loop gains.
The results of our simulations are shown in Fig. \ref{fig:star}\textbf{b}. As can be seen, the energy increases exponentially with the number of target nodes $p$ at a given distance $d$, with the rate of exponential growth depending on the particular choice of the parameter $\delta$, i.e., the smaller the value $\delta$ the higher is the rate.\\
\indent
It is interesting to see that however, the rate of exponential growth of the energy is little affected by the distance of the driver nodes $d$.
This is shown in table \ref{tab:star} as we vary both the distance $d$ and the parameter $\delta$.
\begin{table}[!t]
  \renewcommand{\arraystretch}{1.3}
  \caption{Approximate slopes for the curves in Fig. \ref{fig:star}\textbf{b}}
  \label{tab:star}
  \centering
  \begin{tabular}{lccc}\toprule
     & $d=1$ & $d=2$ & $d=3$\\\midrule
     $\delta=0.1$ & $3.4$ & $3.4$ & $3.6$ \\
     $\delta=0.2$ & $3.0$ & $3.0$ & $3.1$\\
     $\delta=0.5$ & $2.5$ & $2.4$ & $2.7$\\\bottomrule
  \end{tabular}
\end{table}
The observed variations of the exponential rate with the distance $d$ are by far less significant than those seen when the parameter $\delta$ is varied.\\
\indent
From the results in \cite{klickstein2017energy}, the expected value of the energy increases exponentially with the number of target nodes, but the variation of energy values is large as can be seen by the wide standard deviation bars across multiple orders of magnitude.
We consider Erdos-Renyi (ER) graphs with $n=400$ and $\ell$ edges so that the average degree $\frac{2\ell}{n}$.
For every pair of nodes $(v_i,v_j)$ we compute the distance $d_{ij}$ and the energy when $v_i$ is the only driver node and $v_j$ is the only target node, $E_{ij}^*$.
The mean of the energy is plotted with respect to the distance between the two nodes with error bars representing one standard deviation are shown in Fig. \ref{fig:ER}.
Also included is the energy values from a path graph with the same value $k$ as in all of the ER graphs.
We see that the energy along the path graph provides an upper bound to the energy between any two nodes in the ER graphs.
As we have assumed there are no \emph{multi-edges}, there can be at most one path of length $d=1$ between two nodes.
As such, one would expect that the energy requirement when the driver node and target node are adjacenct would be approximately the same as the energy of the path graph when $d = 1$.
We see this is the case in Fig. \ref{fig:ER} for all ER graphs examined when $d = 1$.
When $d = 2$, the possibility of multiple paths of length $2$ arises and we begin to see some energy values are less than the pair of nodes at distance $d=2$ in the path graph.
As $d$ increases from $3$ to $5$, the number of possible paths of length $d$ increases.
The average energy between two nodes of distance $d$ increases more slowly than the path graph as the density of the ER graph grows.
In the most sparse graph, $\ell = 400$, the energy requirements between pairs of nodes is nearly equal to the path graph.\\
\indent
The minimum and maximum energy over any pair of nodes of distance $d$ for the ER graphs is shown in Fig. \ref{fig:max}.
For the case $\ell = 400$ in Fig. \ref{fig:max}\textbf{a}, so that the average degree is $2$, we see very little deviation from the energy values of the path graph.
The average degree of the path graph is $2(n-1)/n \approx 2$ as well.
For the other ER graphs, the maximum value of the energy between a pair of nodes at distance $d$ remains nearly equal to the path graph.
The minimum value of the energy on the other hand remains within a few orders of magnitude of the path graph energy as was predicted by the balloon graph.
\section{Conclusion}
It has been established previously that the control energy required to drive a subset of nodes (the target nodes) of a complex network scales exponentially with the cardinality of the set of target nodes \cite{klickstein2017energy}.
Here, we examine in more detail the role any individual node plays as a target node when paired with a driver node.
Given a distance between driver node and target node $d$, we have seen the path graph of length $d$ where the driver node and target node are on opposite ends provides an upper bound on the energy contribution.
Additional paths of the same length can reduce the energy by a few orders of magnitude.
Including additional targets at the same distance $d$ from the sole driver node exponentially increases the energy but the rate of the increase is independent of the particular distance and instead is a function of the individual self-loop weights, characterized by the parameter $\delta$.
Finally, we examine the energy between pairs of driver nodes and target nodes in complex networks and we see that our predictions based on the path graph, balloon graph and star graph hold.\\
\indent
This work provides the framework for two design aspects of complex networks.
Given a driver node target node pair at distance $d$, one can reduce the energy requirement by either adding an edge to reduce the distance between them $d' < d$, or adding an edge to increase the number of paths of length $d$.
On the other hand, adding an additional edge that accomplishes neither of these tasks will not yield any benefit.
Rather than adding edges, if one is interested in removing edges, any edge that is neither a part of one of the shortest paths between the driver node and target node will not increase the control energy.
In the future, these relations can be used to optimize the design of complex networks with respect to the control energy.
\section*{Acknowledgement}
This work was funded by the National Science Foundation through NSF grant CMMI- 1400193, NSF grant CRISP- 1541148 and ONR Award No. N00014-16-1-2637.
\bibliographystyle{IEEEtran}

\end{document}